\documentclass[fleqn,twoside,psfig]{article}
\usepackage{espcrc2}

\usepackage{graphicx}
\usepackage{epsfig}
\usepackage{axodraw}

\newcommand{\AmS}{{\protect\the\textfont2
  A\kern-.1667em\lower.5ex\hbox{M}\kern-.125emS}}

\def\simgt{\rlap{\lower 3.5 pt\hbox{$\mathchar \sim$}}\raise 1pt \hbox {$>$}}
\def\simlt{\rlap{\lower 3.5 pt\hbox{$\mathchar \sim$}}\raise 1pt \hbox {$<$}}

\newcommand{\kslash}{k\kern-1ex /}
\newcommand{\pslash}{p\kern-1ex /}
\newcommand{\qslash}{q\kern-1ex /}
\newcommand{\lslash}{l\kern-1ex /}
\newcommand{\sslash}{s\kern-1ex /}
\newcommand{\Dslash}{{\cal D}\kern-1.5ex /}

\newcommand{\beqa}{\begin{eqnarray}}
\newcommand{\eeqa}{\end{eqnarray}}

\newcommand{\be}{\begin{equation}}
\newcommand{\ee}{\end{equation}}
\newcommand{\ben}{\begin{eqnarray}}
\newcommand{\een}{\end{eqnarray}}

\def\lsim{\raise0.3ex\hbox{$<$\kern-0.75em\raise-1.1ex\hbox{$\sim$}}}
\def\gsim{\raise0.3ex\hbox{$>$\kern-0.75em\raise-1.1ex\hbox{$\sim$}}}
\def\simgt{\rlap{\lower 3.5 pt\hbox{$\mathchar
\sim$}}\raise 1pt \hbox {$>$}}
\def\simlt{\rlap{\lower 3.5 pt\hbox{$\mathchar
\sim$}}\raise 1pt \hbox {$<$}}
\newcommand{\cont}{{\rm cont}}
\newcommand{\latt}{{\rm latt}}

\newcommand{\csw}{{c_{\rm SW}}}

\newcommand{\ce}{{\it c}_{\it E}}
\newcommand{\cb}{{\it c}_{\it B}}

\newcommand{\pos}{{p^*}}
\newcommand{\qos}{{q^*}}
\newcommand{\lo}{{(0)}}
\newcommand{\nlo}{{(1)}}
\newcommand{\mplo}{{m_p^{(0)}}}

\def\ovec{\partial_\mu\hspace{-0.4cm}\raisebox{1.8ex}{$\rightarrow$}}
\def\antivec{\partial_\mu\hspace{-0.4cm}\raisebox{1.8ex}{$\leftarrow$}}

\title{One-loop calculation of mass dependent ${\cal O}(a)$
improvement coefficients for the relativistic heavy
quarks on the lattice}

\author{Sinya~Aoki$^{\rm a}$,
        Yasuhisa~Kayaba\address{Institute of Physics,
             University of Tsukuba,
             Tsukuba, Ibaraki 305-8571, Japan} and
        Yoshinobu~Kuramashi\address{
             High Energy Accelerator Research Organization(KEK),
             Tsukuba, Ibaraki 305-0801, Japan}}
\begin{document}

\begin{abstract}
We carry out the one-loop calculation of mass dependent 
${\cal O}(a)$ improvement coefficients in 
the relativistic heavy quark action recently proposed, employing 
the ordinary perturbation theory with the fictitious gluon 
mass as an infrared regulator.
We also determine renormalization factors and
improvement coefficients for the axial-vector current at the one-loop level.
It is shown that the improvement coefficients 
are infrared finite at the one-loop level if and only if the improvement
coefficients in the action 
are properly tuned at the tree level.

\end{abstract}

\maketitle

\section{INTRODUCTION}

While, in principle, lattice QCD allows a precise determination of weak
matrix elements associated with the $B$ and $D$ mesons,
$c$- and $b$-quarks are too heavy to treat
directly in current numerical simulations due to large 
${\cal O} (m_Q a)$ errors.

Recently, a new relativistic approach to control $m_Q a$ errors was
proposed from the view point of the on-shell ${\cal O}(a)$
improvement program assuming $m_Q \gg \Lambda_{\rm QCD}$\cite{akt,latt02}.
The action is given by 
\[
S_q= a^4 \mbox{$\sum_x$} {\bar q}(x)\left[ \gamma_0 D_0
+\nu \gamma_i D_i +m_0 \right. 
\]
\[
-a r_t/2 D_0^2-a r_s/2 D_i^2-a ig \cb/4 \sigma_{ij} F_{ij}
\]
\be
\left.
-a ig \ce/2 \sigma_{0i} F_{0i}
\right]q(x),
\label{eq:qaction}
\ee
where we are allowed to choose $r_t=1$ and other four parameters
$\nu$, $r_s$, $c_E$ and $c_B$ are analytic functions of $m_Q a$ and 
the gauge coupling $g$.
Once a nonperturbative adjustment of four parameters
$\nu$, $r_s$, $c_E$ and $c_B$ is achieved,
the remaining cutoff effects are reduced to be 
${\cal O}((a\Lambda_{\rm QCD})^2)$. 
In this report we present a perturbative determination
of $c_E$ and $c_B$ up to the one-loop level
(see Ref.~\cite{latt02} for details).
In addition we calculate the renormalization factors and the ${\cal O}(a)$ 
improvement coefficients for the axial-vector current at the one-loop level.

\section{DETERMINATION OF $\ce$ AND $\cb$ AT THE ONE-LOOP LEVEL}

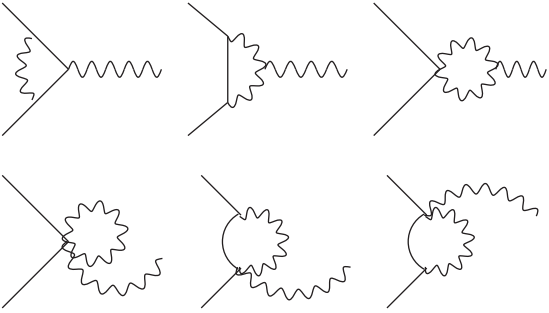
\begin{figure}[t]
\begin{picture}(200,120)(-10,0)

\thicklines
\Line(-10,10)(15,35)
\Line(-10,60)(15,35)
\PhotonArc(25,38)(10.5,-180,180){2}{10}
\PhotonArc(33,35)(18.5,-180,-20){2}{7}
\Line(65,10)(80,25)
\Line(65,60)(80,45)
\CArc(85,35)(12,-240,-120)
\PhotonArc(85,35)(11,-120,120){2}{8}
\PhotonArc(100,42)(27,-142,-38){2}{7}
\Line(135,10)(150,25)
\Line(135,60)(150,45)
\CArc(155,35)(12,-240,-120)
\PhotonArc(155,35)(11,-120,120){2}{8}
\PhotonArc(170,28)(27,38,141){2}{7}
\Line(-10,125)(15,100)
\Line(-10,75)(15,100)
\Photon(15,100)(50,100){3}{5}
\PhotonArc(15,100)(18,-220,-140){2}{4}
\Line(60,125)(75,112.5)
\Line(60,75)(75,87.5)
\Line(75,112.5)(75,87.5)
\PhotonArc(75,100)(12.5,-90,90){2}{6}
\Photon(89,100)(120,100){3}{4}
\Line(130,125)(155,100)
\Line(130,75)(155,100)
\PhotonArc(165,100)(10,-180,180){2}{10}
\Photon(176,100)(195,100){3}{3}

\end{picture}
\vspace{-1cm}
\caption{One-loop quark-gluon vertex diagrams.}
\label{fig:vtx}
\end{figure}

We determine $\ce=\ce^{(0)}+g^2\ce^{(1)}$ and 
$\cb=\cb^{(0)}+g^2\cb^{(1)}$ from 
the on-shell quark-quark scattering amplitude.
In the massless case,  
$\csw$ is successfully determined up to 
the one-loop level employing the conventional
perturbation theory with the fictitious gluon mass as an infrared 
regulator~\cite{c_sw}.
We extend this calculation to the massive case.
Since the tree level determination of $\ce$ and $\cb$ in the massive case 
is already performed in Ref.~\cite{akt}, next step is 
to calculate six types of one-loop diagrams depicted 
in Fig.~\ref{fig:vtx}.

We first consider to calculate $\cb^{(1)}$.
Without the space-time symmetry the general form of the
off-shell vertex function at the one-loop level is written
as
\[
\Lambda_k^{(1)}(p,q,m)\]
\[=\gamma_k F_1^k
+\gamma_k\{\pslash F_2^k+\pslash_0 F_3^k\}
+\{\qslash F_4^k+\qslash_0 F_5^k\}\gamma_k
\]
\[
+\qslash\gamma_k\pslash F_6^k
+\qslash\gamma_k\pslash_0 F_7^k
+\qslash_0\gamma_k\pslash F_8^k
\]
\be
+(p_k+q_k)\left[ H_1^k+\pslash H_2^k
+\qslash H_3^k+\qslash\pslash H_4^k\right]
\label{eq:vtx_k_offsh}
\ee
\[
+(p_k-q_k)\left[ G_1^k+\pslash G_2^k
+\qslash G_3^k+\qslash\pslash G_4^k\right]
+{\cal O}(a^2),
\]
where $\Lambda_k=\Lambda_k^\lo
+g^2\Lambda_k^\nlo+{\cal O}(g^4)$ and
$\pslash=\sum_{\alpha=0}^3 p_\alpha \gamma_\alpha$, 
$\qslash=\sum_{\alpha=0}^3 q_\alpha \gamma_\alpha$, 
$\pslash_0=p_0 \gamma_0$, 
$\qslash_0=q_0 \gamma_0$. 
$p$ and $q$ are incoming and outgoing quark momenta, respectively.
The coefficients $F_i^k$, $G_i^k$ and $H_i^k$ are
functions of $p^2$, $q^2$, $p\cdot q$ and $m$. 
Sandwiching $\Lambda_k^{(1)}(p,q,m)$ by the on-shell
quark states
$u(p)$ and $\bar u(q)$, which satisfy
$\pslash u(p) = i m_p u(p)$ and $\bar u(q) \qslash =
im_p \bar u(q)$,
the matrix element is reduced to
\[
{\bar u}(q)\Lambda_k^{(1)} (p,q,m)u(p)
\]
\[
={\bar u}(q)\gamma_k u(p)
\left\{F_1^k+i m_p 2 F_2^k-m_p^2 F_6^k\right\}
\]
\[
+(p_k+q_k){\bar u}(q)u(p)
\left\{H_1^k+im_p 2 H_2^k-m_p^2 H_4^k\right\}
\]
\be
+{\cal O}(a^2),
\label{eq:vtx_k_onsh}
\ee
where we use the relations among $F_i^k$, $G_i^k$ and $H_i^k$
constrained by the charge conjugation symmetry. 

The relevant term for the determination of $\cb$ is the
last one in eq.(\ref{eq:vtx_k_onsh}), whose coefficient 
can be extracted by setting
$p=\pos\equiv (p_0=im_p, p_i=0)$ and
$q=\qos\equiv (q_0=im_p, q_i=0)$ in
eq.(\ref{eq:vtx_k_offsh}):
\[
\left[H_1^k+im_p (H_2^k+H_3^k)-m_p^2
H_4^k\right]_{p=\pos,q=\qos}^{\latt}
\]
\be
=\frac{1}{8}{\rm Tr}
\left[
\left\{\frac{\partial}{\partial p_k}
+\frac{\partial}{\partial q_k}\right\}
\Lambda_k^\nlo(\pos,\qos,m)
(\gamma_4+1)\right.
\label{eq:cb_from_Lambda}
\ee
\[
\left.
-\left\{\frac{\partial}{\partial p_i}
-\frac{\partial}{\partial q_i}\right\}
\Lambda_k^\nlo(\pos,\qos,m)(\gamma_4+1)\gamma_i\gamma_k
\right]^{i\ne k}.
\]

\begin{figure}[t]
\centering{
\psfig{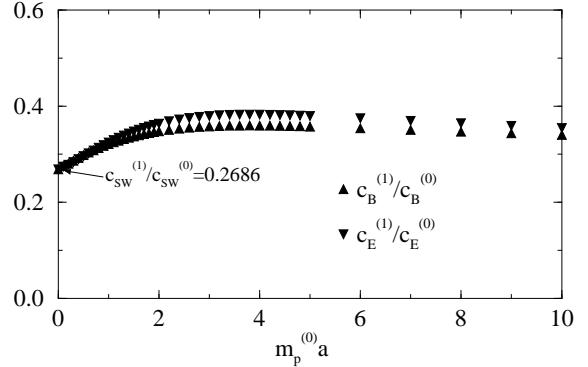}
}
\vspace{-10mm}
\caption{$\ce^{(1)}$ and $\cb^{(1)}$ with the plaquette
gauge action as a function of pole mass $m_{p}^{(0)}$.}
\label{fig:ce_cb}
\end{figure}
We should remark that the second term in
eq.(\ref{eq:vtx_k_onsh})
contains both the lattice artifact of ${\cal O}(p_ka, q_ka)$
and the physical contribution of ${\cal O}(p_k/m,q_k/m)$.
The parameter $\cb$ is determined to eliminate
the lattice artifact of ${\cal O}(p_ka, q_ka)$:
\[
\frac{\cb^{(1)}-r_s^{(1)}}{2}=\left[H_1^k+i m_p 2 H_2^k
-m_p^2 H_4^k\right]_{p=\pos,q=\qos}^\latt
\]
\[
-Z_q^\lo\left[H_1^k+im_p 2 H_2^k
-m_p^2 H_4^k\right]_{p=\pos,q=\qos}^\cont.
\]

The infrared behavior of eq.(\ref{eq:cb_from_Lambda}) is 
investigated by expanding the inner loop momentum 
in Fig.~\ref{fig:vtx} around zero.
The infrared divergence is found to be
\[
\left(-\frac{1}{2N_c}+\frac{N_c}{2}\right)(\cb^\lo-r_s^\lo)L
-\left(-\frac{N_c}{2}\right)\frac{Z_q^\lo}{\mplo}L
\]
where $L=-1/(16\pi^2)\ln|\lambda^2 a^2|$ with $\lambda$ the
fictitious gluon mass.
If the tree level improvement coefficients are properly tuned 
as $\cb^\lo=r_s^\lo$,
we are left with $-(N_c/2)({Z_q^\lo}/{\mplo})L$, which is
exactly the same as the infrared divergence in the
continuum theory with the correct normalization factor $Z_q^{(0)}$.

In the same way, $\ce^{(1)}$ is determined free from the infrared
divergence.
Results for $\cb^{(1)}$ and $\ce^{(1)}$ with the
plaquette action are plotted in Fig.~\ref{fig:ce_cb}.
As expected $\cb^{(1)}$ and $\ce^{(1)}$ approaches to the
massless value for $\csw$\cite{csw_w} as $m_p^{(0)} a$ vanishes.

\section{${\cal O}(a)$ IMPROVEMENT OF THE AXIAL-VECTOR CURRENT}

\begin{figure}[t]
\vspace{-4mm}
\centering{
\psfig{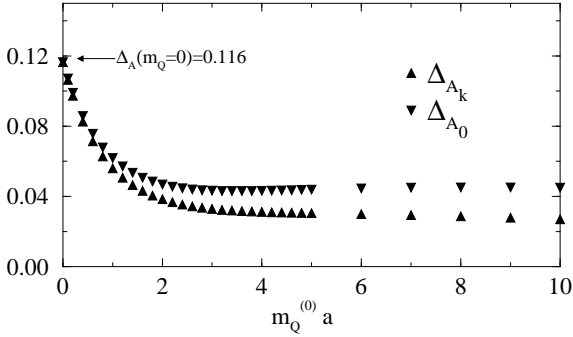}}
\vspace{-10mm}
\caption{$\Delta_{A_\mu}$ in heavy-light system with the plaquette
gauge action as a function of heavy quark pole mass
$m_Q^{(0)}$.}
\label{fig:axhl}
\vspace{-2mm}
\end{figure}

We consider the heavy-light axial-vector current $A_\mu$.
Without space-time symmetry, the renormalized
operator $A^{\latt,R}_\mu(x)$ with the ${\cal O}(a)$ improvement is written as
\[
A^{\latt,R}_\mu(x)=
Z_A^{\latt} \left[ \hspace{0.1cm} 
{\bar q(x)} \gamma_\mu\gamma_5 Q(x) \right.
\]
\[
-g^2 c_{A_\mu}^+
\partial_\mu^+ \{{\bar q(x)} \gamma_5 Q(x)\}
-g^2 c_{A_\mu}^- \partial_\mu^- \{{\bar q(x)}
\gamma_5 Q(x)\}
\]
\[
-g^2 c_{A_\mu}^L \{{\vec
\partial_i}{\bar q(x)}\} \gamma_i 
\gamma_\mu\gamma_5 Q(x) 
\]
\be \left.
-g^2 c_{A_\mu}^H {\bar q(x)}
\gamma_\mu\gamma_5 \gamma_i \{{\vec \partial_i} Q(x)\}
\hspace{0.1cm} \right]
\ee
where $Z_{A_\mu}^{\latt}$ and $c_{A_\mu}^{(+,-,H,L)}$ depend on
the quark masses $m_Q$ and $m_q$. 
$\partial_\mu^+$ and $\partial_\mu^-$ are defined as
$\partial_\mu^+ = \ovec + \antivec$ and 
$\partial_\mu^- = \ovec - \antivec$.
With the aid of equation of motion we can choose
$c^H_{A_0}=c^L_{A_0}=0$.

\begin{figure}[tr]
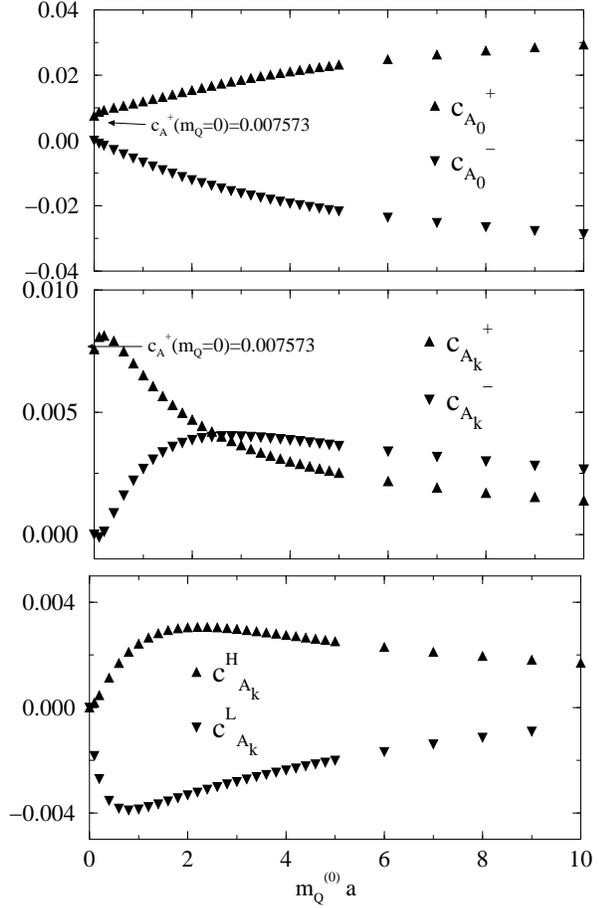

\vspace{-4mm}
\flushright{
\psfig{file=a4yzhl.eps2,height=75mm,angle=-90}}
\flushright{
\vspace{-0.4cm}
\psfig{file=a1yzhl.eps2,height=75mm,angle=-90}}
\flushright{
\vspace{-0.50cm}
\psfig{file=a1whl.eps2,height=76.25mm,angle=-90}}
\vspace{-10mm}
\caption{$c_{A_0}^{(+,-)}$, $c_{A_k}^{(+,-,H,L)}$ in
heavy-light system with the plaquette gauge action}
\label{fig:ayzhl}
\vspace{-2mm}
\end{figure}

To extract $Z_{A_\mu}^{\latt}$ and $c_{A_\mu}^{(+,-,H,L)}$ 
at the one-loop level from the general form of the vertex
function, we have to use both decay and scattering processes.
Except for the physical contribution, all parameters are
infrared finite, once $\nu^{(0)}$,
$r_s^{(0)}$, $\ce^{(0)}$ and $\cb^{(0)}$ are properly
tuned. In Figs.~\ref{fig:axhl},~\ref{fig:ayzhl}, 
we show numerical results for $\Delta_{A_\mu}$
and $c_{A_\mu}^{(+,-,H,L)}$ in heavy-light system. 
The former is defined by
$Z_A^{\latt}/Z_A^{\overline{\rm MS}}=
\sqrt{Z_{Q}^{(0)}}(1-g^2\Delta_{A_\mu})$,
where we take $m_q=0$ for the light quark.
We observe anticipated features 
that $c_{A_\mu}^{(-,H,L)}$ vanishes
as the heavy quark mass decreases, while 
$\Delta_{A_\mu}$ and $c_{A_\mu}^+$ become close 
to their corresponding massless values obtained in Refs.~\cite{csw_w,c_a}. 

This work is supported in part by the Grants-in-Aid for
Scientific Research from the Ministry of Education, 
Culture, Sports, Science and Technology.
(Nos. 13135204, 14046202, 15204015, 15540251, 15740165.)

\end{document}